\documentclass[11pt]{article}
\newcommand{\al}{\alpha}
\newcommand{\bt}{\beta}
\newcommand{\de}{\delta}

\newcommand{\fr}{\frac}
\newcommand{\ga}{\gamma}

\newcommand{\ka}{\kappa}
\newcommand{\La}{\Lambda}
\newcommand{\la}{\lambda}
\newcommand{\lb}{\label}
\newcommand{\lf}{\left}
\newcommand{\Lg}{{\cal L}}
\newcommand{\nn}{\nonumber\\}
\newcommand{\om}{\omega}
\newcommand{\rh}{\rho}
\newcommand{\rt}{\right}
\newcommand{\Si}{\Sigma}
\newcommand{\si}{\sigma}
\newcommand{\sq}{\sqrt}
\newcommand{\ta}{\tau}

\newcommand{\te}{\theta} 
\newcommand{\vb}{\verb}

\newcommand{\be}{\begin{equation}}
\newcommand{\ee}{\end{equation}} 
\newcommand{\eei}{\end{equation}\indent\indent}
\newcommand{\bc}{\begin{center}}
\newcommand{\ec}{\end{center}}
\newcommand{\ber}{\begin{eqnarray}}
\newcommand{\ear}{\end{eqnarray}}
\newcommand{\ba}{\begin{array}}
\newcommand{\ea}{\end{array}}
\newcommand{\p}{\partial}

\def\case#1/#2{\textstyle\frac{#1}{#2} }
\begin{document}
\title{Spacetime Exterior to a Star:\\  Against Asymptotic Flatness.}
\author{Mark D. Roberts,
54 Grantley Avenue,  Wonersh Park GU5 0QN\\
mdrobertsza@yahoo.co.uk
http://cosmology.mth.uct.ac.za/$\sim$roberts} 
\maketitle
{\scriptsize\tableofcontents}
\begin{abstract}
In many circumstances the perfect fluid conservation equations can be directly 
integrated to give a {\it geometric-thermodynamic} equation:  
typically that the lapse $N$ is the reciprocal of the enthalpy $h$, 
($ N=1/h$).   
This result is aesthetically appealing as it depends only 
on the fluid conservation equations 
and does not depend on specific field equations such as Einstein's.   
Here the form of the geometric-thermodynamic 
equation is derived subject to 
spherical symmetry and also for the shift-free ADM formalism.   
There at least three 
applications of the geometric-thermodynamic equation,  
the most important being to the 
notion of asymptotic flatness and hence to spacetime exterior to a star.   
For asymptotic flatness one wants $h\rightarrow 0$ and $N\rightarrow 1$ 
simultaneously,  but this is incompatible with the 
geometric-thermodynamic equation.   
A first shot at modeling spacetime exterior to a star 
is to choose an idealized geometric configuration 
and then seek a vacuum-Einstein solution.   
Now the assumption of a vacuum is an approximation,  
in any physical case there will be both fields and fluids present.   
Here it is shown that the requirement that the star is 
isolated and the presence of a fluid are incompatible in most isentropic 
cases,  the pressure free case being an exception.   
Thus there is the following dilemma:  either an astrophysical system 
cannot be isolated or the exterior fluid must 
depend explicitly on the entropy. 
That a system cannot be isolated is another way of stating Mach's principle.  
The absence of asymptotically flat solutions depends on:  
i)the equation of state,  
ii)the admissibility of vector fields,  
and iii)the requirement that the perfect fluid permeates the whole spacetime.
The result is robust against different choices of 
geometry and field equations because it just depends 
on the fluid conservation equations and 
the ability to introduce a suitable preferred vector field.  
For example with spherical symmetry 
there is the preferred vector field tangent to the 3-sphere;  
furthermore for asymptotically flat 
spacetimes there is the preferred vector field tangent 
to the 3-sphere at infinity.   
The properties of spherically symmetric geodesics are studied both to examine
whether a fluid spheres geodesics can explain solar system dynamics and also
to model hypothetical galactic halo with spherically 
symmetric fluids so as to produce constant galactic rotation curves.
The rate of decay of fields and fluids is discussed in particular whether 
there are any non-asymptotically flat decays and what gravitation modifies 
the Yukawa potential to be.
The Tolman-Ehrenfest relation follows immediately from $N=1/h$.   
The equations relating the enthalpy to 
the lapse have a consequence for the cosmic censorship hypothesis 
this is briefly discussed.
\end{abstract}
\section{Introduction.}\lb{intro}
\subsection{A Relation Between Fluids and Geometry.}\lb{flu}
The perfect fluid stress can be covariantly differentiated 
to give the perfect fluid conservation equations.   
In many cases these differential equations can be directly integrated 
to give a {\it geometric-thermodynamic} equation,  
which typically equates the Eisenhart (1924) \cite{bi:eisenhart}
Synge (1937) \cite{bi:synge} fluid index $w$ 
to the reciprocal of the lapse $N$.   
The Eisenhart-Synge index is essentially the fluids zero temperature enthalpy.
In section \ref{sec:ecs} the index is calculated for several 
equations of state.   
The $\al-$equation of state is a $2-$parameter equation of state which 
describes polytropes.   
The $\bt-$equation of state is a $1-$parameter equation of state obtained 
from the $\al-$equation of state by assuming 
the second law of thermodynamics for an adiabatic process,  
Tooper (1965) \cite{bi:tooper},  Zeldovich and Novikov (1971) \cite{bi:ZN}.   
It gives the $\ga-$equation of state in all cases except for 
$\ga=0$ where the pressure free ($p=0$) case is not recovered;  
but rather  $\mu=p\ln(\fr{p}{K})$.  
For equations of state see also Eligier {\it et al} (1986) \cite{bi:EGH},
and Ehlers \cite{bi:ehlers}.
The index is not defined in the 
pressure free case,  thus solutions such as the 
Tolman (1934) \cite{bi:tolman}-Bondi (1947) \cite{bi:bondi} 
solution are not covered by description in terms of the fluid index.
\subsection{It's Application}\lb{apl}
The main application of the 
geometric-thermodynamic equation is to the description 
of spacetime exterior to a star.   
It is shown in many cases that asymptotic flat solutions do not exist.   
This is taken to imply that the notion of asymptotic flatness 
as usually understood is physically simplistic.   
In the literature diagrams are constructed which are supposed to 
represent the causal spacetime of a collapsing star.   
These diagrams usually require that the 
spacetime is asymptotically flat,  
but the inclusion of a non-vacuum stress is often sufficient for 
this requirement no longer to hold.   
An example of how these diagrams can be qualitatively altered by
infinitesimal matter is given by solutions to the scalar-Einstein
equations which often have no event horizons,  Roberts (1985) \cite{bi:mdr85}.
To the lowest approximation the spacetime exterior to a 
star has no stress:  the star exists in a vacuum.   
In order to take account of the matter that 
surrounds a star it is necessary to find 
an approximate stress which has contributions from 
planets,  dust  etc\dots
There seems to be no systematic way of producing a stress which 
approximates such diverse forms of matter.   
When relativity is applied to macroscopic 
matter the stress is usually taken to be a perfect fluid,  
so that this is taken to be the form of 
the first order correction to the vacuum.   
Specifically the stress is taken to be a spherically 
symmetric perfect fluid with $\ga-$equation of state 
and the result generalized where possible.
The nature of the surface of the star is left open 
as boundary conditions to interior solutions are not 
discussed.   The assumed equations of state are essentially of one variable
so that the pressure $p$ does not depend on the entropy $s$,  
i.e.$~p=p(\mu)$ not $p=p(\mu,s)$ or in other words
they are isentropic.   Stars radiate and the radiation possess entropy 
whether there are equations of state that can describe this,
and if so whether they are susceptible to a similar analysis as to that 
given here is left open:  however it would be unusual for at the simplest 
level there to be asymptotically flat spacetimes,  at the next level of
complexity for there to be none,  and at the full level of complexity for 
asymptotically flat spacetime to reappear.
\subsection{Asymptotic Flatness.}\lb{afl}
It will be shown that many spacetimes with a perfect fluid stress do not have 
asymptotically flat solutions.   
Throughout it is assumed that the fluid permeates the whole 
spacetime and that the spacetime is of infinite extent.  
Also throughout it is assumed that a simple limiting process is appropriate:
specifically as the luminosity radial coordinate tends to infinity the
spacetime becomes Minkowskian.   This does not always happen,  two examples
are: Krasi\'{n}ski's (1983) \cite{bi:krasinski} analysis of the Stephani
universe where $r=\infty$ labels a second center of symmetry,  and type B
Bekenstein conformal scalar extensions of ordinary scalar field solutions,
Agnese and LaCamera (1985) \cite{bi:ALC},  and Roberts (1996) \cite{bi:mdr96},
which have bizarre asymptotic properties.
The result follows from the conservation 
equations so that it is {\bf explicitly} 
independent of the gravitational field equations used.   
The conservation equations use a Christoffel symbol 
or a generalization of this.   The connection 
depends on the metric which in turn can be thought 
of as a solution to gravitational field equations.   
In this {\bf implicit} sense the result can be thought of 
as depending on field equations.   
It might not hold if there are other fields coupled to the fluid.   Similarly 
asymptotically flat solutions are rare for theories with quadratic Lagrangians,
Buchdahl (1973) \cite{bi:buchdahl}.   The absence 
of asymptotically flat solutions might have application 
to the "missing mass" problem,
see Roberts (1991) \cite{bi:mdr91} and references therein.
Some properties of exact perfect fluid solutions have been 
discussed by Delgaty and Lake (1998) \cite{bi:DL}.
In Bradley {\it et al} (1999) \cite{bi:BFMP} it is shown that the Wahlquist 
perfect fluid spacetime cannot be smoothly joined to an exterior
asymptotically flat vacuum region. 
Boundary conditions for isolated horizons are described
in Ashtekar {\it et al} (1999) \cite{bi:ABF}.
Models of stars in general relativity have been discussed by 
Herrera {\it et al} (1984) \cite{herrera},
Stergioulas (1998) \cite{bi:ster}, and Nilsson and Uggla (2000) \cite{bi:NU}.
\subsection{Sectional Contents.}\lb{scn}
Section \ref{sec:ecs} introduces the stress,  conservation equations,  
and the relationship between the enthalpy $h$ and the Eisenhart-Synge 
fluid index $\om$.   
In section \ref{sec:affs} it is shown that 
there are no asymptotically flat static fluid spheres unless 
the fluid index $\om\rightarrow1$ at infinity,  
and using Einstein's field equations there are 
no asymptotically flat static fluid spheres with 
$\ga-$equation of state.   
In the non-static case for $\ga-$equation of state there are no asymptotically 
flat solutions provided that $\ga\ne0,1$ and certain conditions 
hold on the metric.
For both static and non-static cases there might be 
asymptotically flat solutions for $\al-$polytropes.   
In section \ref{sec:gte} it is shown for static spacetimes 
admitting non-rotating vector $U_{\al}=(N,0)$ 
and having $\ga-$equation of state that the lapse $N$ is 
inversely proportional to the fluid index $N=1/\om$.  
For the non-static case, 
subject to $\dot{N}=0$ and $\ga(\ln(\sq{g^{(3)}}))_{,i}=0$,
the equation relating the lapse to the fluid index is 
$\om=N^{-1}g^{(3)\fr{1}{2}(1-\ga)}=\mu^{\fr{\ga-1}{\ga}}$.   
These results can be used to show that there are 
no asymptotically flat fluid filling 
spacetimes admitting the vector $U_{a}=(N,0)$ with $\ga-$equation of state,
again also subject to certain conditions on the metric.
The introduction of the vector $U_{a}=(N,0)$ 
assumes that the fluid is non-rotating and that the spacetime admits a global 
time coordinate,  unlike the vacuum Einstein equations,  
see for example Cantor et al (1976) \cite{bi:CFM},
and Witt (1986) \cite{bi:witt}.   
In section \ref{sec:aaf} a case against asymptotic flatness is presented.
Outer solar system observations of orbital irregularities are discussed.
Non-asymptoticness on length scales greater than the solar system,  such as
galaxies,  is mentioned.   
The time-like geodesics for an arbitrary Newtonian potential are calculated.
Modeling hypothetical galactic halos of ``dark matter'' with spherically
symmetric fluid solutions so as to produce constant galactic rotation 
curves is attempted.   The the rate of decay of various fields are discussed.
It is argued that most perfect fluid spheres and some conformal scalar spheres 
rate of decay is in fact an increase prohibiting asymptotic flatness.  
There is the possibility of experimentally testing gravitational theory by
measuring the deviation of the Yukawa potential from what would be expected 
in the absence of gravitation;  how this might be done is briefly discussed,
the possibility of an actual test seems remote.
Various onion models of spacetime surrounding the Sun are discussed.
It is argued that non-asymptoticness implies that a system cannot be 
gravitationally isolated and that this suggests a new formulation of 
Mach's principle:  {\sc there are no flat regions of physical spacetime.}
The philosophy of what an ``isolated system'' entails is briefly discussed.
In section \ref{sec:ter} the 
Tolman-Ehrefest (1930) \cite{bi:TE} relation is derived.   
Section \ref{sec:cc} speculates on the relevance of the 
geometric-thermodynamic relation to cosmic censorship.
\section{The Enthalpy and the Eisenhart-Synge fluid Index.}
\label{sec:ecs}
\subsection{Perferct Fluids.}\lb{tpf}
The stress of a perfect fluid is given by
\be
T_{\al\bt}=(\mu+p)~U_{\al}U_{\bt}+p~g_{\al\bt}
=nh~U_{\al}U_{\bt}+p~g_{\al\bt},~~~~~~~U_{\al}U_{.}^{\al}=-1,
\ee
where $\mu$ is the fluid density,  $p$ is the pressure,  
$n$ is the particle number,  $h$ is the enthalpy,  
and $p+\mu=nh$.   The unit timelike vector $U_{a}$ 
defines the geometric objects
\ber
h_{\al\bt}&=&g_{\al\bt}+U_{\al}U_{\bt},~~~
\dot{U}_{\al}=U_{\al;\bt}U_{.}^{\bt},~~~
\te=U^{\al}_{.;\al},~~~	
K_{\al\bt}=U_{\chi;\de}h_{\al.}^{~\chi}h_{\bt.}^{~\de},\nonumber\\	
\om_{\al\bt}&=&h_{\al.}^{~\chi}h_{\bt.}^{~\de}U_{[\chi;\de]},~~~
\si_{\al\bt}=U_{(\al;\bt)}+\dot{U}_{(\al}U_{\bt)}-\fr{1}{3}\te h_{\al\bt},
\label{eq:geob}
\ear
called the projection tensor,  the acceleration,  the expansion,  
the second fundamental form,  the rotation,  and the shear,
see for example page 83 of Hawking and Ellis \cite{bi:HE}.
The projection obeys $U_{\al}h^{\al}_{. \bt}=0$ 
and $\dot{U}_{\al}h^{\al}_{. \bt}=\dot{U}_{\bt}$,
also the acceleration obeys $U^{\al}_{.}\dot{U}_{\al}=0$.
Formally the second fundamental form and its associated 
hypersurface only exist when the rotation vanishes.  
Transvecting the stress conservation 
equation $T_{\al.;\bt}^{~\bt}$ with $U_{.}^{\al}$ 
and $h^{\al}_{.\ga}$ gives the first conservation equation
\be
-U^{\al}_{.}T^{\bt}_{\al . ; \bt}=
\mu_{\al}U_{.}^{\al}+(\mu+p)U^{\al}_{.;\al}=
\dot{\mu}+(\mu+p)\te=0
\label{eq:ce1}
\ee
and the second conservation equation
\be
h^{\al}_{. \ga}T^{\bt}_{\al . ;\bt}=
(\mu+p)\dot{U}_{\al}+h_{\al.}^{~\bt}p_{\bt}=0,
\label{eq:ce2}
\ee
respectively.   These equations equate 
the derivatives of the vector field to the pressure and density.   
From a technical point of view,
here we are investigating when these equations can be integrated.   
It turns out that assuming a specific form of vector field - say hypersurface 
orthogonal $U_{\al}=\la\phi_{,\al}$ is not directly of much use,  
but rather assumptions about the form of the metric have to be made.   
The first law of thermodynamics can be taken in the infinitesimal form
\be
dp=n~dh+nT~ds,
\label{eq:1stlaw}
\ee
where $T$ is the temperature and $s$ is the entropy.   
The Eisenhart\cite{bi:eisenhart}-Synge\cite{bi:synge} fluid index is defined by
\be
\ln(\om)\equiv\int\fr{dp}{(\mu+p)}
\label{eq:esfi}
\ee
after setting $T=0$ in \ref{eq:1stlaw} and integrating it is apparent that
up to a constant factor at zero temperature $\om=h$.   The index is also
discussed on page 84 of Hawking and Ellis \cite{bi:HE}.  
\subsection{Polytropes}\lb{ply}
The $\al-$polytrope has equation of state 
\be
p=\al\mu^{\bt}
\label{eq:aeq}
\ee
and has
\be
dp=\al\bt\mu^{\bt-1}d\mu,
\label{eq:daeu}
\ee
or
\be
\fr{\p p}{\p \al}=\fr{\p p}{\p \bt}=0,
\label{eq:paeq}
\ee
because of this the pressure is not an explicit function of two variables
$\al$ and $\bt$,  but only one.   
The index and particle number corresponding to \ref{eq:aeq} are
\be
\om=(1+\al\mu^{\bt-1})^{\fr{\bt}{\bt-1}},~~~~~
n=\mu(1+\al\mu^{\bt-1})^{\fr{1}{1-\bt}},  
\label{eq:apoly}
\ee
The $\bt-$polytrope \cite{bi:ZN} has equation of state
\be
p=Kn^{\ga},
\label{eq:beq}
\ee
where $K$ is a constant and $V=1/n$ is the volume occupied by one bayron.
For an adiabatic process (no exchange of heat) the second law of 
thermodynamics is 
\be
p=-\fr{\p E}{\p V},
\label{eq:2ndlaw}
\ee
where $E$ is the total energy density per unit mass $E=\mu/n$.   
Then \ref{eq:2ndlaw} becomes
\be
p=n^{2}\fr{\p \mu /n}{\p n},
\label{eq:210}
\ee
\ref{eq:beq} and \ref{eq:210} give
\be
pn^{-2}=
K\mu_{o}^{\ga-2}=
\fr{\p \mu/n}{\p n},
\label{eq:211}
\ee
which in the case $\ga\ne1$ can be integrated to give
\be
\mu=\fr{K}{\ga -1}n^{\ga},
\label{eq:int1}
\ee
where the constant of integration is taken to be zero.   
Using \ref{eq:beq},  \ref{eq:int1} becomes the 
equation of state of $\ga-$polytrope
\be
p=(\ga-1)\mu,~~~~\ga\ne 1,
\label{eq:geq}
\ee
which has index and particle number
\be
\om=\mu^{\fr{\ga-1}{\ga}},~~~~~	
n=\ga\mu^{\fr{1}{\ga}},~~~~~
\ga\ne0,1,
\label{eq:gpoly}
\ee
In the pressure free case ($\ga=1$ in \ref{eq:geq})
the index \ref{eq:esfi} is not defined,  an option is to
replace $p$ with $(\ga-1)\mu$ in the definition \ref{eq:esfi}
and then take $\ga=1$ to obtain $\ln(\om)=0$ or $\om=1$,
then the condition $n\om=\mu+p$ gives $n=\mu$.
For the $\ga-$equation of state the first \ref{eq:ce1} and second \ref{eq:ce2}
conservation laws can be written in terms of $\mu$,  
where $\mu=\om^{\fr{\ga}{\ga-1}}$,  and are 
\be
\dot{\mu}+\ga\mu\te=0,
\label{eq:gac1}
\ee
and
\be
\ga\mu\dot{U}_\al+(\ga-1)h_{\al.}^{~\bt}\mu_\bt=0,
\lb{eq:gac2}
\ee
respectively.
The $\ga-$equation of state has been derived under 
the assumption that $\ga\ne1$.   
Perhaps the correct $\ga=1$ equation of state for a $\bt-$polytrope 
is found by putting $\ga=1$ in \ref{eq:211} and integrating to give
\be
\mu=p\ln\left(\fr{p}{K}\right);
\label{eq:int2}
\ee
however the speed of sound 
\be
v_s\equiv\fr{\p p}{\p \mu}=\left(\ln\left(\fr{p}{K}\right)+1\right)^{-1},
\label{eq:speedsound}
\ee
is $1$ or the speed of light when $p/K=1$,  it is less than the speed of 
light for $p/K>1$,  and it diverges as $p/K\rightarrow\exp(-1)$.   
That the speed of sound can take these values 
suggests that this equation of state is essentially 
non-relativistic.   Some writers refer to \ref{eq:int2} as dust,  
others call the pressure free case $p=0$ dust.
\ref{eq:int2} has index and particle number
\be
\om=\left(1+\ln(\fr{p}{K})\right)^{\fr{1}{K}},~~~~~
n=p\left(1+\ln(\fr{p}{K})\right)^{\fr{K-1}{K}}.
\label{eq:216}
\ee
\section{Asymptotically Flat Fluid Spheres}
\label{sec:affs}
\subsection{Spherical Symmetry.}\lb{ssy}
The line element of a spherically symmetric spacetime can be put in the form
\be
ds^{2}=-C~dt^{2}+A~dr^{2}+B~d\Si^{2}.
\label{eq:ssst}
\ee
Choosing the timelike vector field
\be
U_{a}=(\sqrt{C},0,0,0),
\label{eq:tlvf}
\ee
the rotation vanishes and the projection tensor,  acceleration,  
expansion,  shear,  and second fundamental form are
\ber
h_{r.}^{~r}&=&h_{\te.}^{~\te}=h_{\phi.}^{~\phi}=1,\nn
\dot{U}_{\al}&=&(0,\fr{C'}{2C},0,0),\nn
\te&=&-\fr{1}{\sqrt{C}}(\fr{\dot{A}}{2A}+\fr{\dot{B}}{B}),\nn
\si_{rr}&=&-\fr{2A}{B}\si_{\te \te}=-\fr{2A}{Bsin^{2}\te}\si_{\phi\phi}
=\fr{1}{3}\fr{1}{\sqrt{C}}(-\dot{A}+\fr{A}{B}\dot{B}),\nn
K_{rr}&=&-\fr{1}{2}\fr{1}{\sqrt{C}}\dot{A},\nn
K_{\te \te}&=&\fr{1}{sin \te}K_{\phi \phi}=-\fr{1}{2}\fr{1}{\sqrt{C}}\dot{B},
\lb{eq:L1}
\ear
where the overdot denotes absolute derivative with respect to $\ta$ 
as in $\dot{U}_{\al}=\fr{D U_{\al}}{d\ta}$,  
but otherwise the overdot denote partial derivative with respect to time.   
Noting that $d\mu/d\tau=dt/d\tau~\mu_{,t}=\dot{\mu}/\sqrt{C}$,
the first conservation equation \ref{eq:ce1} becomes
\be
\dot{\mu}-(\mu+p)(\fr{\dot{A}}{2A}+\fr{\dot{B}}{B})=0,
\label{eq:34a}
\ee
and the second conservation equation \ref{eq:ce2} becomes
\be
p'+(\mu+p)\fr{C'}{2C}=0,
\label{eq:34b}
\ee
only the $r$ component is non-vanishing in the second equation.
\subsection{The Static Case.}\lb{sca}
In the static case the first conservation equation \ref{eq:34a}
vanishes identically and the second conservation equation 
\ref{eq:34b} integrates to give
\be
\om=\fr{1}{\sqrt{C}},
\label{eq:com}
\ee
the constant of integration is taken to be independent of $\te$ and $\phi$ 
and is absorbed into $C$,  for example by redefining $t$.   
For the line element \ref{eq:ssst} to be asymptotically flat it is 
necessary that as $r\rightarrow\infty$, the line element \ref{eq:ssst} 
becomes Minkowski spacetime in other words as r increases 
$C\rightarrow 1$,   $A\rightarrow 1$  and $B\rightarrow r^{2}$.  
Now from \ref{eq:com},  $C\rightarrow 1$ implies that $\om\rightarrow 1$.
Thus any static spherical fluid sphere with a well defined index not equal 
to $0$ or $1$ cannot be asymptotically flat.    
To see this result in particular cases first consider the $\ga-$equation of 
state.	From \ref{eq:gpoly} and \ref{eq:com}
\be
\mu=C^{\fr{\ga}{2(1-\ga)}},
\label{eq:37}
\ee
and as $C\rightarrow 1$,  $\mu$ tends to a constant 
and thus the spacetime cannot be asymptotically flat;  
also the spacetime cannot be asymptotically DeSitter 
as this would necessitate $\mu$ tending to a constant time $r^{2}$.  
In the pressure free case,  the index is not defined and there are the 
asymptotically flat solutions given by Tolman \cite{bi:tolman} 
and Bondi \cite{bi:bondi}.   
Next consider the $\bt-$equation of state,   from \ref{eq:216} and \ref{eq:com}
\be
C=\left(1+\ln(\fr{p}{K})\right)^{-\fr{2}{K}},
\label{eq:38}
\ee
now asymptotically as $C\rightarrow 1$,  $p\rightarrow K$;
however a constant value of $p$ asymptotically is not 
consistent with asymptotic flatness,
therefore there are no asymptotically flat solutions.
Finally consider the $\al-$equation of state,  
from \ref{eq:apoly} and \ref{eq:com}
\be
C=(1+\al\mu^{\bt-1})^{\fr{2\bt}{1-\bt}},
\label{eq:39}
\ee
in the case $\mu\rightarrow 0$, $C\rightarrow 1$ and there 
might be asymptotically flat $\al-$polytropic spheres.   
The same results are obtained using the more general vector
\be
U_{\al}=(a\sqrt{C},\sqrt{(a^{2}-1)A},0,0),
\label{eq:310}
\ee
where $a$ is a constant.
\subsection{The Non-static Case.}\lb{nns}
In the non-static case it is necessary 
to assume an equation of state in order to 
calculate a geometric-thermodynamic relation.
The $\ga-$equation of state is assumed.
Then either from \ref{eq:gac1} and \ref{eq:gac2} and \ref{eq:L1},
or from \ref{eq:geq} and \ref{eq:34a} and \ref{eq:34b} the first and
second conservation laws are
\be
\dot{\mu}-\ga\mu\left(\fr{\dot{A}}{2A}
                                       +\fr{\dot{B}}{B}\right)=0,
\lb{eq:A1}
\ee
and
\be
(\ga-1)\mu'+\ga\mu\fr{C'}{2C}=0,
\lb{eq:A2}
\ee
respectively.   The equation
\ber
d\mu&=&\dot{\mu}dt+\mu'dr\nn
    &=&-\ga\mu\left(\fr{\dot{A}}{2A}+\fr{\dot{B}}{B}\right)dt
       +\fr{\ga\mu}{\ga-1}\fr{C'}{2C}dr,
\lb{eq:A3}
\ear
can be integrated when
\be
\dot{C}=0,
\lb{eq:A4}
\ee
and
\be
(AB^2)'=0,
\lb{eq:A5}
\ee
to give
\be
\mu=\om^{\fr{\ga}{\ga-1}}=A^{+\fr{\ga}{2}}B^{+\ga}C^{\fr{\ga}{2(1-\ga)}},~~~
\ga\ne0,1,
\label{eq:311}
\ee
where the constant of integration have been taken to be 
independent of $\te$ and $\phi$ and is 
absorbed into the line element.   
The assumption $(AB^2)'=0$ is coordinate dependent and holds rarely as 
for example $(AB^2)'\approx 4r^3$ for Minkowski spacetime in spherical
coordinates whereas $(AB^2)'=0$ 
for Minkowski spacetime in rectilinear spacetime.
Taking the limits $A,C\rightarrow 1$,  
$B\rightarrow r^{2}$,  for $\ga>0$,  $\mu \rightarrow$ a constant, 
and for $\ga<0$, $\mu$ diverges;   
thus there are no asymptotically flat solutions.  The $\al-$equation of 
state \ref{eq:aeq} cannot be investigated without further information.
Discussion of non-existence of time dependent fluid spheres can also be found 
in Mansouri (1977) \cite{bi:mansouri}.
\section{The Geometric-Thermodynamic equation in the ADM formalism.}
\label{sec:gte}
\subsection{Vanishing Shift ADM Formalism.}\lb{vsf}
In the ADM (-1,+3) \cite{bi:ADM} formalism with vanishing shift the metric 
is given by
\be
g_{\al\bt}=(-N^{2},g_{ij}),~~~
g^{\al\bt}=(-N^{-2},g^{ij}),~~~
\sqrt{-g^{(4)}}=N\sqrt{g^{(3)}}.
\label{eq:41}
\ee 
where $g^{(3)}$ is the determinant of the $3-$dimensional metric.   
The reason the shift is taken to 
vanish will become apparent later.   The timelike unit vector field used here
\ber
U_{\al}&=&(N,0),~~~~~~
U^{\al}=(-\fr{1}{N},0),\nn
U_{i;t}&=&-N_{,i},~~~~~~
U_{i;j}=-\fr{1}{2N}g^{(3)}_{ij,t},\nn
U_{t;t}&=&U_{t;i}=0,
\label{eq:42}
\ear
there are other choices such as $U_{\al}=(-N,0)$,
and also $U_{\al}=(aN,bN_{i})$ 
for which the unit size condition $U_{\al}U^{\al}_{.}=-1$
implies $g^{ij}N_{i}N_{j}=\fr{a^{2}-1}{b^{2}}$.   
For \ref{eq:42} the rotation vanishes and the remaining 
geometric objects \ref{eq:geob} are
\ber
h_{ij}&=&g_{ij},~~~~~~
\dot{U}_{\al}=(0,\fr{N_{i}}{N}),~~~~~~
\te=-\fr{1}{N}\left(\ln(g^{(3)})\right)_{,t},\nn
\si_{ij}&=&-g^{(3)}_{ij,t}
           +g^{(3)}_{ij}\left(\ln(g^{(3)})\right)_{,t},\nn 
K_{ij}&=&K_{ji}=-g_{ij,t}.
\label{eq:43}
\ear
The first conservation equation \ref{eq:ce1} becomes
\be
\mu_{,t}-(\mu+p)\left(\ln\sqrt{g^{(3)}}\right)_{,t}=0,
\label{eq:44a}
\ee
and the second conservation equation \ref{eq:ce2} becomes
\be
p_{i}+(\mu+p)\fr{N_{,i}}{N}=0,
\label{eq:44b}
\ee
the $t$ component of the second conservation equation 
\ref{eq:44b} vanishes identically.   
If  the shift is included in the above vector \ref{eq:42} one finds
\be
2N^{2}U^{0}_{.i}=2NN_{,i}+(N_{k}N^{k})_{,i}
                +N^{j}(2N_{j,i}-N^{k}g_{\vb+{+ik,j\vb+}+}),
\label{eq:withshift}
\ee
and further calculation proves intractable.
\subsection{Static Case.}\lb{stc}
In the static case the first conservation equation vanishes identically 
and the second conservation equation integrates immediately 
and independently of the equation of state to give
\be
\om=\fr{1}{N},	
\label{eq:45}
\ee
where the constant of integration has been absorbed into $N$.   
\subsection{Non-static Case.}\lb{nsc}
In the non-static case assume the $\ga-$equation 
of state has to be assumed to accommodate 
the first conservation law \ref{eq:ce1}.          
With $\ga-$equation of state \ref{eq:geq} the conservation equations 
\ref{eq:44a} and \ref{eq:44b} integrate to give
\be
\om=\fr{1}{N}g^{(3)\fr{1}{2}(\ga-1)},~~~~~~~\ga\ne0,1,
\label{eq:46}
\ee
where in place of \ref{eq:A4} and \ref{eq:A5}
\be
\dot{N}=0,
\lb{eq:A6}
\ee
and
\be
\ga\left(\ln\left(\sqrt{g^{(3)}}\right)\right)_{,i}=0,
\lb{eq:A7}
\ee
respectively.   Constants of integration have been absorbed into the line 
element.   Substituting the spherically symmetric values of the previous 
section into \ref{eq:46} gives \ref{eq:311} times a function 
of $\sin\te$ which has been taken to be absorb able there.  
The equations \ref{eq:45} and \ref{eq:46}
depend on the choice of velocity vector \ref{eq:42},  
for example if a  geodesic velocity vector is chosen then the 
acceleration vanishes and \ref{eq:45} and \ref{eq:46} do not hold.    
The conditions \ref{eq:A6} and \ref{eq:A7} do not appear to have an invariant
formulation.   There are three things to note.   The {\it first} is that these
derivatives do not occur in the covariant derivatives of the vector field
\ref{eq:42} and hence do not occur in the geometric objects \ref{eq:43}.
The {\it second} is that \ref{eq:A4} and \ref{eq:A7} are satisfied if
\be
\{^t_{tt}\}=0,
\lb{eq:A8}
\ee
and
\be
\{^i_{jk}\}=0,
\lb{eq:A9}
\ee
respectively,  as they only occur in these Christoffel symbols.
The {\it third} is that \ref{eq:A6} and \ref{eq:A7} might {\bf solely} be a
gauge condition;  but \ref{eq:A6} puts on one constraint and \ref{eq:A7} puts
on three  constraints totaling four, the usual number of differential
gauge constraints.
The Plebanski-Ryten (1961) \cite{bi:PR} gauge condition is
\be
[(-g)^wg^{ab}_{..}]_{,b}=0,
\lb{eq:A10}
\ee
for $w=\fr{1}{2}$ this is the harmonic gauge condition.
For $a=t$,   \ref{eq:A10} is
\be
-\fr{1}{N^2}(\ln(g^{(3)w}))_{,t}+\fr{\dot{N}}{N^3}=0.
\lb{eq:A11}
\ee
For $a=x^i$,   \ref{eq:A10} is
\be
-\fr{N_{,i}}{N}+(\ln(g^{(3)w})g^{ij}_{..})_{,j}=0.
\lb{eq:A12}
\ee
For $w\ne0$,   \ref{eq:A6} and \ref{eq:A7} cannot be recovered,
except for Minkowski spacetime in rectilinear coordinates.  
Thus the conditions \ref{eq:A6} and \ref{eq:A7} on the metric appear not to be
an example of Plebanski-Ryten gauge conditions.
It can be asked,  is there a non-static geometric-thermodynamic relation 
which involves familiar gauge conditions instead of metric constraints such 
as \ref{eq:A6} and \ref{eq:A7}.   Inspection of \ref{eq:gac1} and \ref{eq:gac2}
with arbitrary vector field instead of \ref{eq:42} does not immediately give
a choice of vector field for which application of the Plebanski-Ryten gauge
\ref{eq:A10} simplifies matters enough for the problem to be tractable.
\subsection{$\ga$-equation of state and the ADM.}\lb{gad}
For the $\ga-$equation of state \ref{eq:46} becomes
\be
\mu=N^{\fr{\ga}{1-\ga}},~~~~~~~~~~~~\ga\ne0,1,	
\label{eq:48}
\ee
for the spacetime to be asymptotically flat the density $\mu$ must vanish
asymptotically implying that the lapse $N$ must vanish,  
contradicting the assumption that the spacetime is asymptotically flat.   
For the $\ga-$equation of state \ref{eq:geq},  \ref{eq:45} becomes
\be
\mu=N^{\fr{\ga}{1-\ga}}g^{(3)\ga/2},~~~~~\ga\ne0,1,
\label{eq:49}
\ee
asymptotically $\mu\rightarrow r^{2}$ 
and the spacetime cannot be asymptotically flat.  
For $\al-$polytropes the static case \ref{eq:45} gives
\be
N=\left(1+\al\mu^{\bt-1}\right)^{\fr{\bt}{1-\bt}},
\label{eq:410}
\ee
and in this case it is possible for $N\rightarrow 1$ and 
$\mu\rightarrow 0$ simultaneously as $r \rightarrow\infty$
Thus for spacetimes where the rotation free vector 
\ref{eq:42} can be introduced,  and subject to the caveats mentioned above
for the non-static case:  
i)there are no asymptotically flat $\ga-$polytropes except possibly for 
$\ga=0$ or $1$, 
ii)there are no asymptotically flat fluid spacetimes unless the fluid 
index tends to a finite non-vanishing constant.   
\section{Against Asymptotic Flatness.}
\lb{sec:aaf}
\subsection{Lenght Scales.}\lb{lsc}
On length scales from the outer solar system to cosmology there are 
observations indicating that asymptotic flatness of the systems under 
consideration are not correct.   It is known that the dynamics of the outer
solar system have unexplained irregularities.   For example from the figures 
of Seidelmann {\it et al} (1980) \cite{bi:SKPSV} it appears that the 
irregularity in Pluto's orbit is that the RA increases by about 2 arcsec more
than expected in 50 years,  similarly the declination decreases by about 
1 arcsec in 50 years.   The irregularities are not neatly expressible by a
single quantity,  as for example the orbit of Mercury was prior to general
relativity;  but roughly this means that the orbit is boosted by about 
2 arcsec in 50 years.   This makes the construction of theories to explain the 
irregularities difficult.   In Roberts (1987) \cite{bi:mdr87} the effect 
of a non-zero cosmological constant was investigated in order to explain the 
irregularities of Pluto's orbit and it was found that 
the cosmological constant would have to be about 12 orders of magnitude 
bigger than the upper bound Zel'dovich (1968) \cite{bi:zeldovich} finds
from cosmological considerations.   
Axenides {\it et al} (2000) \cite{AFP} 
also discuss dynamical effects of the cosmological constant.

Scheffer (2001) \cite{scheffer} and Anderson {\it et al} (2001) 
discuss dynamical irregularities in the paths of spacecraft.
The orbit of comets,  Marsden (1985)
\cite{bi:marsden},  Rickman (1988) \cite{bi:rickman},  and Sitarski (1994)
\cite{bi:sitarski} have unexplained irregularities,  for example
at least 6 comets have unexplained forces acting toward the elliptic.
Qualitatively this is exactly what would be expected from Kerr geodesics
\cite{bi:chandrasekhar} page 363,
\begin{quote}
In summary then,  the bound and the marginally bound orbits must 
necessarily cross the equatorial plane and oscillate about it.
\end{quote}
but qualitatively the effect is many orders of
magnitude out:  on solar system length scales the Kerr modification of 
Schwarzschild geometry is intrinsically short ranged.   
These solar system orbital problems might 
originate from the oblateness of the sun,  Landgraf (1992) \cite{bi:landgraf}.
There are theories which have gravitational potential with an exponential term 
and mass scale $m_p(m_H/m_P)^n$,  where $m_H$ is a typical hadron mass,  
$m_P$ is the Planck mass,  and $n=0,1,$ and sometimes $2$.   
Satellite and geophysical data for $n=2$ theories show that 
they are not viable unless $m_H>10^3GeV$,  
Gibbons and Whiting (1981) \cite{bi:GW}.
Other searches for an adjusted potential have been undertaken by 
Jarvis (1990) \cite{bi:jarvis}.
\subsection{The Exterior Schwarzschild Solution as a Model.}\lb{esm}
The exterior Schwarzschild solution is a reasonable model 
of the solar system outside the sun.  
A fluid solution can be argued to be a better approximation 
to the matter distribution as it takes some account of interplanetary space 
not being a vacuum.   
Any exterior fluid spacetime would have different geodesics than 
the vacuum Schwarzschild solution, 
consequently the orbits of the planets would be 
different from that suggested by the Schwarzschild solution:  
how to calculate these geodesics for spherically symmetric spacetimes
is shown below.   The magnitude of the upper limit of the effective 
cosmological constant is about $\rho_{\La}=10^{-16}{\rm g.~ cm.}^{-3}$,
that this is too small to explain Pluto's irregular orbit was shown in 
Roberts (1987) \cite{bi:mdr87}.  
Thus to explain Pluto's irregular orbit using a fluid
the critical density must be larger than $\rho_{\La}$.   
$\rho_{\La}$ is much larger than the mean density of 
interplanetary space which is of the order of $10^{-29}{\rm g.~ cm.}^{-3}$ 
(or $10^{-5}$ protons ${\rm cm.}^{-3}$).   
The density of interplanetary matter is insignificant compared to the density 
contribution from the planets,  for example for Jupiter 
$\rho_{{\rm Jupiter}}=\fr{3}{4\pi}M_{{\rm Jupiter}}r_{{\rm Jupiter}}^{-3}
\approx 2.10^{-4} {\rm g.~ cm.}^{-3}$,  where the radius 
$r_{{\rm Jupiter}}$ is the semi-major axis of the planets orbit.   
This density is above $\rho_{\La}$ and might be above $\rho_{C}$.   
Taking a fluid to model the planets is an unusual step,  
but the alternative of seeking an $n-$body solution 
to the field equations is not viable because even the $2-$body 
solution is not known.   Looking at constant galactic rotation curves 
one might try an approximation.   As noted in the last paragraph 
of section 5 of \cite{bi:mdr91}:
\begin{quote}
For constant circular velocities over a large distance it is necessary to have
an approximately logarithmic potential.   Thus the metric will have an 
approximately logarithmic term.   The Riemann tensor is constructed from the 
second derivatives of the metric and the square of the first derivatives
of the metric.   For a logarithmic potential these will both be of the 
order $r^{-2}$ and thus a linear analysis might not be appropriate.
\end{quote}
This suggests that only an approach using an exact solution will work.
One can assume that the system under 
consideration can be modeled by a 
static spherically symmetric spacetime with line element \ref{eq:ssst}.
Constructing the geodesics using Chandrasekhar's (1983) \cite{bi:chandrasekhar}
method,  the geodesic Lagrangian is given by
\be
2\Lg=-C\dot{t}^2+A\dot{r}^2+B\dot{\te}^2+B\sin^2\te\dot{\phi}^2.
\lb{eq:gl}
\ee
The momenta are given by
\be
p_a=\fr{\p \Lg}{\p \dot{x}_a}
\lb{eq:mom}
\ee
and are 
\be
p_t=-C\dot{t},~~~
p_r=A\dot{r},~~~
p_\te=B\dot{\te},~~~
p_\phi=B\sin^2\te\dot{\phi}.
\lb{eq:mar}
\ee
Euler's equations are
\be
\dot{p}_a=\p_a\Lg
\lb{eq:Euler}
\ee
For static spacetimes with $\p_t A=\p_t B=\p_tC=0$,  
giving $\fr{\p\Lg}{\p t}=0$ so that the time component of the Euler equation
\ref{eq:Euler} gives $\fr{d p_t}{d \tau}=0$,  integrating
\be
-p_t=C\fr{d t}{d \tau}=E {\rm ~~a~ constant~ along~ each~ geodesic}.
\lb{eq:energy}
\ee
Similarly by spherical symmetry one can take $\p_\phi A=\p_\phi B=\p_\phi C=0$,
giving $\fr{\p \Lg}{\p \phi}=0$ so that the $\phi$ component of the Euler 
equation \ref{eq:Euler} gives $\fr{d p_\phi}{d \tau}=0$,  integrating
\be
p_\phi=B\sin^2\te\fr{d \phi}{d \tau}= {\rm a~ constant}.
\lb{eq:C86}
\ee
For the $\te$ component
\be
\fr{\p\Lg}{\p\te}=B\sin\te\cos\te\fr{d \phi}{d \tau},
\lb{eq:thcom}
\ee
the Euler equation \ref{eq:Euler} is
\be
\fr{d}{d\tau}p_\te=\fr{d}{d\tau}B\dot{\te}
=\fr{\p\Lg}{\p\te}=B\sin\te\cos\te\fr{d\phi}{d\tau},
\lb{eq:theq}
\ee
choosing to assign the value $\pi/2$ to $\te$ when $\dot{\te}$ is zero,
then $\ddot{\te}$ will also be zero;  and $\te$ will remain constant 
at the assigned value.   The geodesic is described in an invariant plane which 
can be taken to be $\te=\pi/2$.   Equation \ref{eq:C86} now gives
\be
p_\phi=B\dot{\phi}=L {\rm ~a~ constant~ along~ each~ geodesic}
\lb{eq:Leq}
\ee
where $L$ is the angular momentum about an axis normal to the invariant plane.
Substituting into the Lagrangian
\be
-\fr{E^2}{C}+A\dot{r}^2+\fr{L^2}{B}=2\Lg=-1 {\rm ~or~} 0,
\lb{eq:sub}
\ee
where $2\Lg=-1{\rm ~or~}0$ depending on whether time-like or null geodesics 
are being considered.   Rearranging
\be
A\dot{r}^2=-\fr{L^2}{B}+\fr{E^2}{C}+2\Lg.
\lb{eq:rea}
\ee
Taking $r$ to be a function of $\phi$ instead of $\tau$ 
and using \ref{eq:Leq} gives
\be
\lf(\fr{dr}{d\phi}\rt)^2=-\fr{B}{A}+\fr{B^2}{AL^2}\lf(\fr{E^2}{C}+2\Lg\rt),
\lb{eq:itp}
\ee
now letting
\be
u\equiv \fr{1}{r}
\lb{eq:defu}
\ee
as in the usual Newtonian analysis
\be
\lf(\fr{du}{d\phi}\rt)^2=-\fr{u^4B}{A}
                         +\fr{u^4B^2}{AL^2}\lf(\fr{E^2}{C}+2\Lg\rt),
\lb{eq:gencase}
\ee
seeking a thermodynamic interpretation one can substitute the enthalpy $h$ for
the lapse $C=h^2$,  but $A$ and $B$ are still arbitrary so that this is not
pursued.   Inserting the K\"ottler (Schwarzschild solution with cosmological 
constant) values of the metric
\be
B=r^2,~~~
C=\fr{1}{A}=1-\fr{2m}{r}+\fr{\La}{3}R^2,
\lb{eq:Kv}
\ee
and taking $2\Lg=-1$ for time-like geodesics equation \ref{eq:gencase} becomes
\be
\lf(\fr{du}{d\phi}\rt)^2=-u^2+2mu^3+\fr{2mu}{L^2}
                   -\fr{1-E^2}{L^2}-\fr{\La}{3u^2L^2}-\fr{\La}{3}
\lb{eq:geoK}
\ee
which is equation (4) of Reference \cite{bi:mdr87},
the last term suggesting the possibility of constant rotation curves.
One can if investigate if there is any adjustment of the Newtonian potential
which will give constant geodesics,  as required for galactic rotation.
Taking (c.f. Will (1993)\cite{bi:will} eq.4.6)
\be
g_{tt}\approx -1+2U,
\lb{eq:n2}
\ee
where $U$ is the Newtonian gravitational potential.   
Now in assume additionally the particular form for 
a spherically symmetric spacetime
\be
B=r^2,~~~
A=\fr{1}{C}\approx \fr{1}{1-2U}\approx 1+2U
\lb{eq:furt}
\ee
inserting in \ref{eq:gencase} and expanding for small $U$ everywhere
\be
\lf(\fr{du}{d\phi}\rt)^2=-\fr{1-E^2}{L^2}-u^2(1+2U)+\fr{2U}{L^2}
\lb{eq:expan}
\ee
In particular one might expect that constant rotation is given by the 
middle term so that
\be
-u^2(1+2U)=\al {\rm ~~a ~constant},
\lb{eq:midal}
\ee
rearranging for $U$ we find
\be
U=-\fr{1}{2}-\fr{\al^2}{2}r^2
\lb{eq:res}
\ee
this suggests that the correct addition to $U$ to produce constant curves
is a function in $r^2$,  this is given by the addition of a cosmological 
constant and such a spacetime is given by K\"ottler's solution \ref{eq:Kv}. 
One might ask what is the next simplest space-time after on with a 
cosmological constant which has an $r^2$ increasing potential and
perhaps this is the interior Schwarzschild solution.
This can be thought of as modeling the halo of a galaxy with the interior
Schwarzschild solution and calculating the geodesics to see if they 
give constant motion.   Newtonian modeling has been done by 
Binney and Tremaine (1987) \cite{bi:BT}.
For the interior Schwarzschild solution Adler,  Bazin,  and Schiffer,
(1975) \cite{bi:ABS} equation number 14.47 one has
\ber
A&=&\fr{1}{1-\fr{r^2}{\hat{R}^2}},~~~
B=r^2,\nn
C&=&\lf[\fr{3}{2}\sq{1-\fr{r_0^2}{\hat{R}^2}}
     -\fr{1}{2}\sq{1-\fr{r^2}{\hat{R}^2}}\rt]^2,\nn
&&{\rm for}~~~
r\le r_0,~~~
\hat{R}^2=\fr{3c^2}{8\pi\ka\rh},
\lb{eq:Schex}
\ear
inserting into \ref{eq:gencase} one gets
\be
\lf(\fr{du}{d\phi}\rt)^2=u^2-\fr{1}{\hat{R}^2}
+\fr{1}{L^2}\lf[-1+\fr{1}{u^2\hat{R}^2}
+\lf(\fr{2E}{3\sq{\fr{u^2\hat{R}^2-u^2r^2_0}{u^2\hat{R}^2-1}}-1}\rt)^2\rt].
\lb{eq:rotis}
\ee
The $\fr{1}{\hat{R}^2}=\fr{8\pi\ka\rh}{3c^2}$term can be thought of as giving
constant rotation curves proportional to the halo density $v_c\propto \rh$.
What one expects from the Tully-Fisher (1977) \cite{bi:TF} relationship
is that $v_c^4\propto L\propto M$.   There might be an exact solution in
Delgaty and Lake (1998) \cite{bi:DL} which will model this more closely. 
\subsection{Rates of Decay.}\lb{rod}
In general one can ask what exact solutions to gravitational field 
equations give what rate of decay.   This problem could also be 
studied numerically.   The rate of decay of scalar fields has been 
discussed in the last paragraph of the introduction of \cite{bi:mdr96}.
Including other fields one roughly gets the rates of decay:
type B conformal scalars $>$ perfect fluids $>$ the gravitational field
$>$ type O scalars $>$ electromagnetic fields $>$ type A conformal scalars 
$>$ coupled and interacting fields and fluids.
The type B conformal scalars and perfect fluids are not usually asymptotically 
flat.   Of course,  for example,  one would expect there to be conformal 
scalar solutions which are neither type A or B and these have unpredictable 
rates of decay,  so this ordering is not absolute.
\subsection{The Spacetime of Elementary Particles.}\lb{els}
Rates of decay are not only important on long distance scales.
As pointed out in the second paragraph of the introduction of \cite{bi:mdr96}
the exact solution of the spherically symmetric spacetime of the 
Klein-Gordon-Einstein equations is not known,  except in the massless case
where the static spherically symmetric field equations $R_{ab}=2\phi_a\phi_b$
have the solution
\ber
ds^2&=&\exp\left(-\fr{2m}{r}\right)dt^2
       -\fr{\eta^4}{r^4}\exp\left(\fr{2m}{r}\right)
          {\rm cosech}^4\left(\fr{\eta}{r}\right)dr^2\nonumber\\
    &&-\eta^2\exp\left(\fr{2m}{r}\right){\rm cosech}^2
                                \left(\fr{\eta}{r}\right)d^2\Sigma,
    ~~~\phi=\sigma/r
\label{eq:4}
\ear
where $\eta^2=m^2+\sigma^2$ and $m$ is interpreted as the mass and $\sigma$ 
the scalar charge,  see for example,  Roberts (1985) \cite{bi:mdr85}.
Only the massless exact solution is known so that the exact modification of
the shape of the Yukawa potential for the meson is not known:
it proves difficult to approximate.   The spacetime of mesons has been 
discussed by Fisher (1948) \cite{bi:fisher},  Ross (1972) \cite{bi:ross},  
Nagy (1979) \cite{bi:nagy},  and Ho (1995) \cite{bi:ho}.   
The Yukawa potential was invented,  
Yukawa (1935) \cite{bi:yukawa},  Landau (1990) \cite{bi:landau},
to account for the $\pi$-meson as the exchange quantum in the force between
two nucleons;  this is by analogy with electromagnetism where it is the 
exchange of a photon that is the origin of the electric and magnetic forces 
between electrons.   The exact form of the potential is 
$V=-(1/r)\exp(-r/m_\pi)$ times a function 
which involves the relative spin orientations.   The Yukawa potential is only
an approximation as Quantum Chromodynamics is really the theory of strong
interaction;  the $\pi$-meson or pion successfully describes the residual 
force,  and is thought to work up to momentum transfer of about 1 GeV/c.
In strong interaction there is also evidence of a linear confining potential.
The Yukawa potential for mesons can be measured using form factors,  several 
mesons are needed,  see for example Gross,  Van Orden and Holinde (1992) 
\cite{bi:VH}.   The hypothetical Higgs particle also has a Yukawa potential.
In order to measure the Higg's Yukawa potential one needs to measure the 
coefficients of the $H^3$ and $H^4$ terms in
\be
V=M_{H^2/2}H^2+\lf(M_{H^2/2v}\rt)H^3+\lf(M_{H^2/2v^2}\rt)H^4,
\lb{eq:sallydawson}
\ee
and verify that the coefficients are as expected.   At present there are no
measurements of these terms.  To measure them one would need to observe 
multi-Higgs production:  this is further discussed in Djouadi {\it et al}
(1999) \cite{bi:DKMZ}.  In the standard model the Higgs coupling enters by
a quartic coupling:  however in supersymmetric theories the quartic couplings 
are connected to gauge couplings which are known,  so that in supersymmetric 
models it is easier to calculate the coefficients.
\subsection{What Happens on Long Distance Scales.}\lb{lds}
If asymptotic flatness is incorrect then what does happen on long scales?
Non-asymptotic flatness introduces the problem of what happens as the potential
goes to infinity - does it increase for ever?
What could happen is that the one body problem becomes inappropriate:
one needs a solution which takes into account two or more bodies.
In particular for the solar system if there is a growing term 
in the potential one might take that it has stopped growing well 
before the next star.   
Torbett and Smoluchowski (1984) \cite{TS} argue that there are bodies 
orbiting the Sun at $10^5~AU\approx5\times10^{-3}pc.$,
which might be a maximum orbiting distance.
Puyoo and Jaffel (1998) \cite{PJ} study the interface between the heliopause
and the interstellar medium,  this is at about $10^3~AU$ and they find a
high interstellar hydrogen density of $0.24\pm0.05~{\rm g.cm.^{-3}}$,
a proton density of $0.043\pm0.005~{\rm g.cm.^{-3}}$,
a helium density of $(2.7\pm0.5)10^{-2}~{\rm g.cm.^{-3}}$, and so forth.
One consequence of these non-vanishing densities is that in gravitation,  
as in quantum field theory,  it becomes difficult to say what a vacuum is 
and whether it has energy, see Roberts (2000) \cite{mdr2000}.
One can ask what sort of metric describes spacetime at various distances
from the Sun,  and it seems that some sort of onion model is called for.
The standard picture is that the interior Schwarzschild solution is matched
to the exterior Schwarzschild solution as in Adler {\it et al} (1975) \S 14.2.,
and then match the exterior Schwarzschild solution to a Friedman model with
a specific equation of state as in Stephani (1985) \cite{bi:stephani} \S 27.3.
Perhaps there should be more than three regions.
The sun has a mean density of $\rh_{Sun}=1.409 {\rm g. cm.^{-3}}$ Allen (1962)
\cite{bi:allen};  however its density varies considerably depending on 
its distance from the centre from Allen (1962) \cite{bi:allen} 
table on page 163;
there is a big jump at about half its radius,  which can be modeled by a dense
core,  so perhaps two interior solutions are needed to describe it.
Dziembowski {\it et al} (1990) \cite{dz} 
and Basu {\it et al} (2000) \cite{basu},
use inversion techniques to show that the sun has many layers with 
different speeds of sound and densities.
The solar system splits up into three regions:
the inner where the general relativistic corrections to 
Newtonian theory are needed,  the middle where Newtonian theory works,
and the outer where a term explaining the irregularity in Pluto's orbit
is needed.   Next one needs a metric to describe the effect of local stars,
then of the galaxy,  and then of groups of galaxies.   the Robertson-Walker
cosmological region comes next,  and after this perhaps a chaotic region.
One can ask if a particle,  say at $1$ parsec from the Sun is not in a
flat region what is it that causes the most deviation from flatness.
For simplicity assume that a Newtonian potential will give correct ratios
between the contributions,  so that the quantity $\phi/G=M/R$ is calculated
in units of the Sun's mass over parsecs.   
A parsec from the Sun is about as isolated as a particle in the nearby galaxy
could be expected to be.
The deviation from flatness of the
metric is approximately given by equation \ref{eq:n2} with $U=\phi$.   
The quantities in
Allen (1962) \cite{bi:allen} \S132,133,135,136,  for the masses and distances
associated with the local star system (Gould belt),  the galaxy,  the local
group of galaxies,  the Universe are used.
Working to the nearest order of magnitude,  the local star system has diameter
$1,000$ pc. and mass $1\times 10^8~M_{Sun}$,  assuming the Sun is near the edge
gives the potential $M/R\approx10^5~M_{Sun}{\rm pc.}^{-1}$.
The galaxy has diameter $25$ kpc. and mass $1.1\times 10^{11}~M_{Sun}$,
but the distance of the Sun from the centre is $8.2\pm0.8$ kpc.,
using this distance $M/R\approx10^8~M_{Sun}{\rm pc.}^{-1}$.
The local group of galaxies consists of $16$ galaxies,  suggesting an 
approximate mass of $10^{12}~M_{Sun}$,  whose centre is $0.4$ Mpc. away 
giving $M/R\approx10^7~M_{Sun}{\rm pc.}^{-1}$.
Van den Berg (1999) \cite{berg} finds $35$ local group members 
and mass $M_{LG}=(2.3\pm0.6)\times10^{12}M_{Sun}$;
and that the zero surface velocity,  which separates the local group
from the field that is expanding with the Hubble flow,
has radius $R_0=1.18\pm0.15~Mpc.$.
The Universe has a characteristic length sale $R=c/H\approx3,000~{\rm Mpc.}$ 
and the mass of the observable Universe is $10^{54}g.$,  again one can form a
ratio $M/R$,  but it has no direct meaning because of homogeneity,  
one finds $M/R\approx 10^{11}$.   To compare with the potential on the surface
of the Earth note that the Earth's mean radius 
$R_{Earth}\approx6\times10^3{\rm Km.}=2\times10^{-12}{\rm pc.}$ and has mass
$M_{Earth}\approx6\times10^{27}{\rm g.}=3\times10^{-6}M_{Sun}$,  giving
$M/R\approx10^6~M_{Sun}{\rm pc.}^{-1}$ for the contribution from the Earth's
mass,  $M/R\approx10^5~M_{Sun}{\rm pc.}^{-1}$ for the contribution from the
Sun's mass.   Collecting these results together gives the ratios
\be
10^6~~:~~10^5~~:~~1~~:~~10^5~~:~~10^8~~:~~10^7~~:~~10^{11}
\lb{eq:ratios}
\ee  
This suggests that either the 
Newtonian approximation is not appropriate,  that asymptotic flatness is not
a physical notion,  or both.
For the onion model it suggests that the metric describing the effect of 
local stars,  the galaxy,  and the local group of galaxies might not be needed
because of the Universe's higher ratio.
Another approach to what sort of notions are useful in describing stellar
systems is as follows.   {\it A priori} one would not wish to exclude the 
possibility that near the centre of the galaxy there are stellar systems 
consisting of several stars,  many planets,  many asteroids and comets,  lots
of dust,  and which are close say only a light year away from other stellar
systems.   Dynamics for such a stellar system perhaps could still be calculated
in some regions,  but there are no notions of a one-body system,  vacuum
field equations,  or asymptotic flatness to use in an explicit manner.
\subsection{Mach's Principle.}\lb{mhp}
Mach's principle can be formulated in many ways:  Barbour and Pfister 
(1995) \cite{bi:BP} p.530 list 21,  Bondi and Samuel (1997) \cite{bi:BS}
list 10.   Different formulations can lead to contradictory conclusions:
for example,  Bondi and Samuel's (1997) \cite{bi:BS} Mach3 and Mach10 give
rise to diametrically opposite predictions when applied to the Lense-Thiring
effect.   A Newtonian formulation has equations which can be used to describe 
dynamics rather than recourse to dark matter,  
Roberts (1985) \cite{bi:mdr85/2}.   Lack of asymptotic flatness suggests that 
a system cannot be isolated.   This is unlike thermodynamics where isolated 
heat baths are ubiquitous,  and unlike electrostatics where the charge inside
a charged cavity can be zero.   So why should a Minkowski cavity in a 
Robertson-Walker universe be excluded?   Field equations and junction 
conditions allow this to be done,  it has to be excluded by principle.
The answer is that it is different 
from electrostatics as gravitation is monopolar in nature.   Any departure 
from homogeneity in the exterior region to a charged cavity would mean a 
change in charge which would quickly attract the opposite charge and cancel 
out:  however in the gravitational case this does not happen,  a change in 
homogeneity exterior to a Minkowski cavity (I think) would quickly change 
the spacetime from being flat.   The above suggests a new formulation of 
Mach's principle:  {\sc there are no flat regions of physical spacetime.}   
What happens for an initial value formulation of this is unclear:  
presumably it means that a 
well-defined initial surface does not develop into a surface part of which 
is flat.   The above statement of Mach's principle is a particular case
of the statement of Einstein (1953) \cite{bi:einstein},  Ehlers (1995) 
\cite{bi:ehlers95},  and Bondi and Samuel (1997) \cite{bi:BS} Mach9  
{\sc there are no absolute elements}:  a flat metric is an absolute element.
\subsection{Isolated Systems.}\lb{iss}
Another way of looking at asymptotic flatness is to note that it implies that 
the solar system is isolated.   {\bf Isolated systems} seem to be an ideal
which is appealed to in order to make problems soluble.   The necessity of
addressing soluble problems is discussed in Medewar (1982) \cite{bi:medawar}.
In practice an isolated system is only an approximation,  there is always
some interaction with the external world and for the assumption of an 
isolated system to work this must be negligible.   The assumption that
systems can be isolated appears through out science,  but there appears to
be no discussion of what this involves in texts in the philosophy of
science.   Three examples of isolated systems are now given.
The {\it first} is photosynthesis:  one can think of each leaf 
on a tree as an isolated entity with various chemical reactions happening
independent of the external world,  but this is only an approximation
as the leaf exchanges chemicals with the rest of the tree so perhaps 
the tree should be thought of as the isolated system,  further one 
can think of the entire biosphere as an isolated entity which converts
$3\times 10^{21}$ Joules per year into biomass from a total of
$3\times 10^{24}$ Joules per year of solar energy falling on the 
Earth,  see for example Borisov (1979) \S 1.2.1 \cite{bi:borisov}.
The {\it second} is in thermodynamics and statistical mechanics,
here the isolatabilty of systems is taken as a primitive undefined concept,
see for example Rosser (1982) \cite{bi:rosser} page 38.
The {\it third} is of experiments where a single electron is taken to be
isolated,  Ekstrom and Wineland (1980) \cite{bi:EW}:  the single electron
is confined for weeks at a time in a ``trap'' formed out of electric and
magnetic fields. 
\section{The Tolman-Ehernfest Relation.}
\label{sec:ter}   
\subsection{The Radiation Fluid.}\lb{trf}
For a radiation fluid $\ga=\fr{4}{3}$,  
and  by \ref{eq:gpoly} the fluid index is
\be
\om=(3p)^{\fr{1}{4}}.
\label{eq:51}
\ee
The Stefan-Boltzman law is
\be
p=\fr{a}{3}T^{4},
\label{eq:52}
\ee
where $T$ is the temperature and $a$ is the Stefan-Boltzmann constant.   
Thus
\be
\om=a^{\fr{1}{4}}T.
\label{eq:53}
\ee
Assuming the spacetime is static 
and admits the rotation free vector \ref{eq:42},  
equations \ref{eq:45} and \ref{eq:53} give
\be
N=a^{-\fr{1}{4}}T^{-1},
\label{eq:54}
\ee
thus showing that the lapse $N$ 
is inversely proportional to the temperature $T$.   
This is the Tolman-Ehrenfest (1930) \cite{bi:TE} relation.    
Lapse only spacetimes have been studied by Roberts (1994) \cite{bi:mdr94}
and Schmidt (1996) \cite{bi:schmidt}.
For the non-static case \ref{eq:46} and \ref{eq:53} gives
\be
Ng^{(3)\fr{1}{6}}=a^{-\fr{1}{4}}T^{-1}.
\label{eq:55}
\ee
\section{The Geometric-thermodynamic equation and Cosmic Censorship.}
\label{sec:cc}
\subsection{Scalar Field Solutuions.}\lb{sfs}
It is known that spherically symmetric asymptotically flat solutions 
to the Einstein massless scalar field equations do not posses event horizons,
both in the static case Roberts (1985) \cite{bi:mdr85} 
and in the non-static case Roberts (1996) \cite{bi:mdr96}.   
Massless scalar field solutions are equivalent to perfect fluid 
solutions with $\ga=2$ and 
$U_{a}=\phi_{a}(-\phi_{c}\phi_{.}^{c})^{-\fr{1}{2}}$;
for the above scalar field solutions the vector field is not 
necessarily timelike so that the perfect fluid correspondence
does not follow through.  It can 
be argued that an asymptotically flat fluid would be 
a more realistic model of a collapsed object,  
because a fluid provides a better representation 
of the stress outside the object.  
In the spherically symmetric case a global coordinate system 
of the form \ref{eq:ssst} can be chosen 
and a necessary condition for there to be an event horizon is that,  
at a finite non-zero value of $r$,  $C\rightarrow\infty$.   
From \ref{eq:com},  \ref{eq:37},  \ref{eq:38},  and \ref{eq:311}  
it is apparent that this only occurs from some exceptional 
equations of state and values for the fluid density.  
Relaxing the requirement of spherical 
symmetry equations \ref{eq:49} and \ref{eq:410} 
show that for there to be a null surface $N\rightarrow 0$,  
or $\om\rightarrow\infty$;  
however  the derivation of both \ref{eq:49} and \ref{eq:410} 
requires the vector \ref{eq:42} and components of this 
diverge as $N\rightarrow 0$,  
also to show that \ref{eq:49} and \ref{eq:410} 
hold globally it is necessary to show that the 
coordinate system \ref{eq:41} can be set up globally.   
The above suggests that it is unlikely that 
spacetimes with a perfect fluid present have 
event horizons except in contrived 
circumstances.
\section{Acknowledgements.}
I would like to thank 
B.G.Masden for discussion about cometary orbits and
Warren Buck,
Sally Dawson,
Helmut Eberl,
Franz Gross.
Paul Stoler,
Erik Woolgar, 
and Peter Zerwas
for discussion about the Yukawa potential.

\end{document}